# Graded negative index lens with designable focal length by phononic crystal

Ke Deng [1,2], Yiqun Ding [1], Zhaojian He [1], Zhengyou Liu [1,3], Heping Zhao [2] and Jing Shi [1]

[1]Key Lab of Acoustic and Photonic materials and devices of Ministry of Education and Department of Physics, Wuhan University, Wuhan 430072, China

[2]Department of Physics, Jishou University, Jishou 416000, Hunan, China

E-mail: zyliu@whu.edu.cn

**Abstract**

We realize a flat lens with graded negative refractive index by a two-dimensional phononic crystal. The index-grade is achieved by gradual modification of the filling fraction along the transverse direction to propagation. We demonstrate that this lens can realize the focusing and amplification of parallel incident acoustic waves. A formula for the refractive index profile is derived by which the lens with desired focal length can be precisely designed and the aberrations of image can be effectively reduced. This designable lens is expected to bear significance in applications such as coupling or integration with various types of acoustic devices.

PACS numbers: 43.20.+g, 43.35.+d, 43.40.+s

   The ability to manipulate waves has been greatly enhanced with the recent advances in metamaterials (MMs) due to the many unusual properties they possess [1]. One of these predominant properties is the negative refraction which has been utilized to image with flat MM lenses [2]. Alternatively, negative refraction can also be realized by photonic/phononic crystals (PCs) which behave like materials with negative refractive index within some frequency regions [3,4]. This feature has been used to design and fabricate flat PC lenses [5-7]. Compared with a MM lens, a PC lens, taking the advantage of much simpler fabrication process, is more suitable for the coupling or integration to other conventional PC devices.

---

[3]Author to whom any correspondence should be addressed.

Unlike a conventional lens made of normal materials, a flat lens made of MMs or PCs does not possess a focal length. It can image a point source in the vicinity but can not focus and amplify the parallel incident waves generated from a distant object. To overcome this shortage, negative index lenses with curved surfaces [8,9] or prism structures [10] have been proposed, but these introduce additional complexity in fabrication process. The graded index lens provides an alternative solution [11-20]. It is well known that a flat slab of optical material with a graded refractive index in the direction transverse to the propagation can focus a parallel beam of light [11]. Now such a concept has been extended to the graded PCs in which the graded index can be achieved more flexibly by appropriate gradual modifications of the lattice constant [20-23], the filling fraction [15-19], or the material parameters [15,16], and so on. These provide us even more degrees in wave manipulations. In the past few years, graded photonic crystals have been extensively explored for the self-collimation [16], focusing [16-20], mirage [21-23], and superbending [22] of electromagnetic (EM) waves.

Recently, a graded lens constructed by a two-dimensional phononic crystal with cylinders arranged in a solid matrix was proposed [15]. This lens works in the first dispersion bands of the PCs and behaves like a media with positive graded index. It was reported that such a lens can focus parallel elastic waves, however, only the shear vertical mode polarized along the cylinders was considered, which constrains its application. In this paper, we propose a flat lens of graded phononic crystal (GPCs) in a fluid matrix to manipulate acoustic waves. Different from [15], our GPC lens works in the second dispersion bands and behaves like an effective media with negative graded index. As demonstrated by the rigid multiple scattering theory (MST), our GPC lens can realize the focusing and amplification of parallel incident acoustic waves at a well defined focal spot. Moreover, we derive a formula for the refractive index profile by which the construct of GPC lens with arbitrary desired focal length can be conveniently accomplished. Unlike [15], in which a prior assignment of the index-profile function was required to derive the focal length [15], here we input a desired focal length then the required index-profile can be derived by the formula and as a consequence, the design of lens is achieved. The aberrations in GPC imaging are also found to be effectively reduced through this design process.

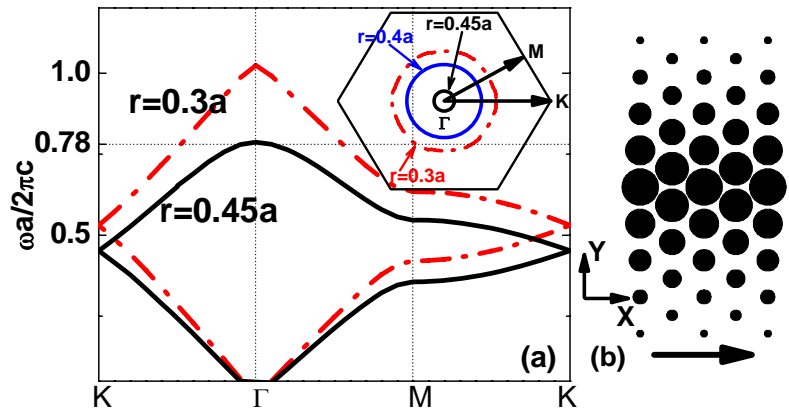

**Figure 1.** (Color online) (a) Band structures for the PCs with $r = 0.3a$ (red dash dot line) and $r = 0.45a$ (black solid line). The inset gives EFSs for the PCs with $r = 0.3a$, $r = 0.4a$ and $r = 0.45a$ at the frequency $0.78$; (b) A schematic illustration for the GPC lens with graded cylinder radii. The arrow below indicates the direction of wave propagation.

The PCs we considered here consist of hexagonal lattice of steel cylinders in air. Our previous work [7] has demonstrated that such PCs possess well defined effective negative refractive index within certain frequency range. Figure 1(a) gives band structures of such PCs with different cylinder radii $r$, which are calculated by the MST [7,24,25]. The material parameters used in the calculations are density $\rho = 1.29\,kg/m^3$ and sound velocity $c = 340\,m/s$ for air;

$\rho = 7.67 \times 10^3 \, kg/m^3$, $c_l = 6010 \, m/s$ and $c_t = 3230 \, m/s$ for steel. The solid and dash dot lines in figure 1(a) represent PCs with $r = 0.3a$ and $r = 0.45a$, respectively, where $a$ is the lattice constant. The inset of figure 1(a) shows the equifrequency surfaces (EFSs) for PCs with different $r$ at normalized frequency $\omega a/2\pi c = 0.78$. As one can infer from figure 1(a), at frequency $0.78$, each PC with $r$ between $0.3a$ and $0.45a$ can be characterized by an effective refractive index $n = -|\vec{k}|c/\omega$ [7]. In addition, the PC with smaller $r$ has a larger modulus of $n$. Therefore, if we construct a slab structure as schematically illustrated in figure 1(b), which consists of cylinders with gradually decreased radius from the center towards the edges along the transverse direction of wave propagation, it can be considered as a negative index material with its refractive index profile in the transverse direction varies in such a way that the modulus of the $n$ is smallest at the center and increases to the edges.

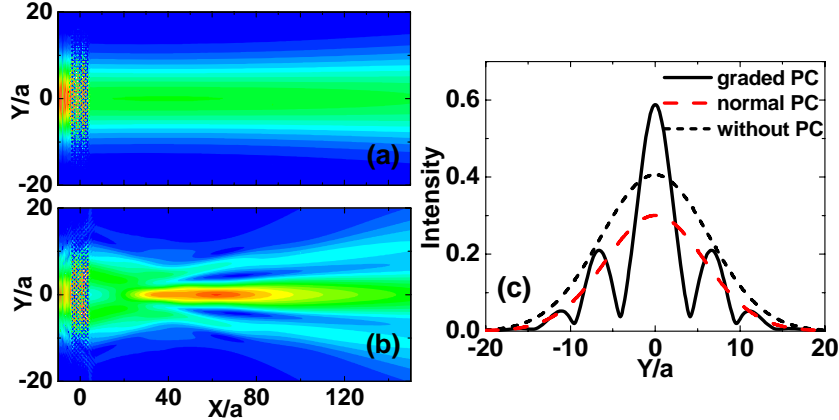

**Figure 2.** (Color online) (a) The intensity distribution of pressure field for a normal PC with no grade which is vertically illuminated by a Gaussian beam; (b) The intensity distribution of pressure field for a linear GPC illuminated by the same Gaussian beam as that in (a); (b) The transverse intensity distribution of pressure field at the focal plane for the case without PC (black short dash line), the normal PC with no grade (red dash line) in (a), and the linear GPC (black solid line) in (b).

Such a slab can be used to focus parallel incident waves [11-14,19]. When a plane wave is incident on it, an off-axis ray will experience a larger phase change than the on-axis ray due to its particular profile of $n$. Since the slab has negative graded indexes, the phase of the off-axis ray becomes ahead of the on-axis ray. This results a convex wave front, thus the waves diverge in the slab. As the consequent of negative refraction, these divergent beams converge at a focal spot behind the slab. This simplified analysis, although applies strictly to a thin slab, still gives a clear illustration for the focusing mechanism of the negative GPC lens. This focusing process is different from the positive GPC lens of Ref. 15, for which the injected parallel waves firstly converged at a focal spot *in the lens* and then became parallel to the direction of propagation again. In order to obtain a real image outside the lens, the slab should be truncated along the direction of propagation before the waves reach that focal spot [15]. Here, in our case, the focal spot is always *outside the lens*. To verify the above analysis, we construct a GPC in which the grade of $n$ is introduced by linearly decreasing of $r$ from $0.45a$ at the centre to $0.3a$ at the edges. We test the focus ability of this linear GPC by vertically illuminating it with a Gaussian beam at frequency $0.78$. A finite system MST [24] is employed to calculate the pressure field which is shown in figure 2. For the purpose of comparison, we firstly calculate the intensity distribution of pressure field for a normal PC with no grade illuminated by the same Gaussian beam. As shown in figure 2(a), there is no focus effect. When the grade is introduced, obvious focus effect can be observed as shown in figure 2(b). Figure 2(c) gives the comparison of transverse intensity distribution of pressure field at the focal plane for the case without PC, the normal PC with no grade, and the linear GPC. We can see that the focusing and magnifying effects of our GPC lens are well manifested in figure 2.

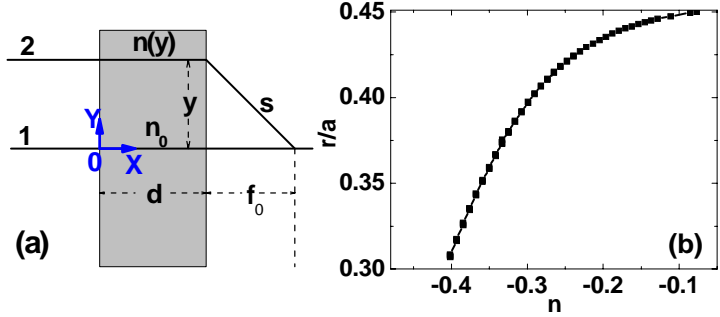

**Figure 3.** (Color online) (a) Schematic drawing for two parallel waves labeled $1$ and $2$ traveling across a slab of materials with a negative graded index $n(y)$; (b) Values of $n$ for PCs with different $r$ at frequency $0.78$.

The linear grade we introduced above is still too simplified for the design of focal length and the reduction of aberrations in GPC imaging. We now derive a formula for the refractive index profile by which the focal length of the GPC lens can be designed and the aberrations can be effectively reduced. We consider two waves labeled $1$ and $2$ which travel across the GPC lens and converge at a focal length $f_0$ behind the slab, as schematically shown in figure 3(a). The wave path is $n_0 d + f_0$ for wave $1$ and $n(y)d + s$ for wave $2$, where $n_0 < 0$ is the refractive index on the $x$ axis and $n(y)$ is the index profile. These two paths must satisfy: $n_0 d + f_0 = n(y) d + s$. Note that $s = \sqrt{y^2 + f_0^{\,2}}$ thus we get the formula of $n(y)$ required by a desired $f_0$:

$$n(y) = n_0 - \frac{1}{d}\left(\sqrt{y^2 + f_0^{\,2}} - f_0\right). \tag{1}$$

Next we calculate by MST the values of $n$ for PCs with different $r$ at frequency $0.78$ as plotted in figure 3(b). Now with equation (1), we can derive the refractive index profile $n(y)$ required for any desired $f_0$. The $n(y)$ can be further transformed to the required radius profile $r(y)$ by the data in figure 3(b). Therefore, the design of lens with desired focal length can be conveniently achieved with the appropriately selection of $r(y)$ in fabrication. As an example, a GPC lens with the desired focal length $f_0 = 88a$ is designed and its focusing effects to a vertically illuminated Gaussian beam are calculated by the finite system MST. figure 4(a) gives the intensity distribution of pressure field. The obtained focal length is $f = 87.8a$ which is in good agreement with the desired one. Figure 4(b) gives the normalized field distribution along the $x$ axis. Figure 4(c) gives the normalized field distribution at the focal plane indicated by figure 4(a). The black (dark) lines in figure 4(b) and (c) are for the designed GPC of figure 4(a). For comparison, the red (gray) lines in figure 4(b) and (c) represent a linear GPC which is otherwise identical to the designed one of figure 4(a). As we can see from figure 4(b) and (c), aberrations in the GPC imaging are also effectively reduced with the designing process. It should be noted that we have simply assumed the wave paths in the lens perpendicular to the slab when deriving the equation (1). Thus our design formula is valid for thin slab and moderate index-grade. The relative error $|f_0 - f|/f_0$ for different $f_0$ are calculated. With the decreasing of $f_0$, the required index-grade becomes less moderate, and thus the relative error $|f_0 - f|/f_0$ increases as shown in figure 4(d).

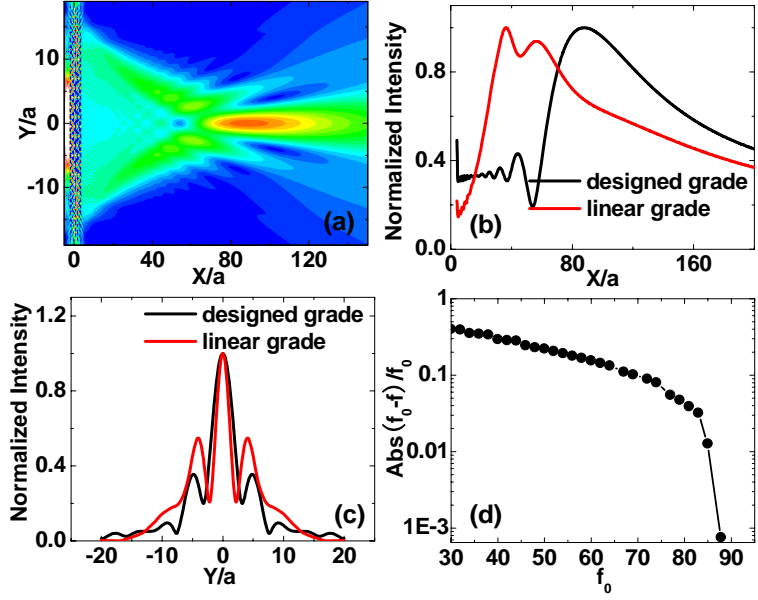

**Figure 4.** (Color online) (a) The intensity distribution of pressure field for a designed GPC which is vertically illuminated by a Gaussian beam. The desired focal length is $f_0 = 88a$ and the obtained one is $f = 87.8a$; (b) The normalized field distribution along the $x$ axis. The black (dark) line is for the designed GPC of (a). The red (gray) line is for a linear GPC which is otherwise identical to the designed one of (a); (c) The normalized field distribution at the focal plane. The black (dark) and the red (gray) lines are the same as those in (b); (d) The relative error $|f_0 - f|/f_0$ as a function of $f_0$.

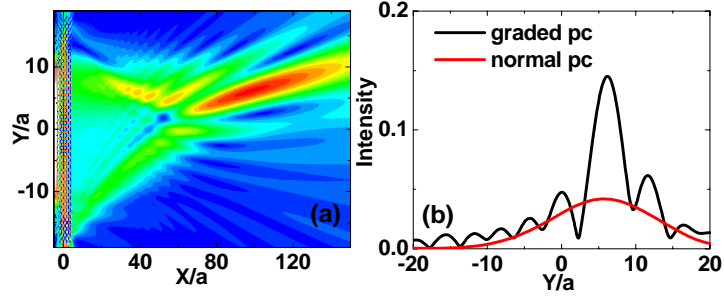

**Figure 5.** (Color online) (a) The intensity distribution of pressure field for the GPC of Fig 4(a) illuminated by a Gaussian beam with incident angle $\theta = 5°$; (b) Transverse intensity distribution of pressure field at the focal plane for the GPC of (a) and for a normal PC with no grade.

The focusing property of our GPC lens is not limited to the vertical incidence. In figure 5 we show the case for the designed GPC of figure 4 illuminated by a Gaussian beam with the incident angle $\theta = 5°$. We can find that the beam is still well focused and magnified for oblique incidence, although the intensity at the focal plane are greatly decreased, which is caused by the rapidly increased reflection.

In conclusion, we have proposed a flat lens of graded phononic crystal to realize the focusing and amplification of parallel acoustic waves. A formula of refractive index profile was derived by which the design of GPC lens with arbitrary desired focal length can be conveniently achieved and the aberrations in the GPC imaging can be effectively reduced. This designable lens is expected to bear significance in applications such as coupling or integration with various types of acoustic devices.

**Acknowledgments**

This work is supported by the National Natural Science Foundation of China (Grant Nos.